\documentclass[journal]{IEEEtran}
\usepackage{amsmath,amsfonts}
\usepackage{algorithmic}
\usepackage{algorithm}
\usepackage{array}
\usepackage{kantlipsum}     
\usepackage{lipsum}
\usepackage{textcomp}
\usepackage{stfloats}
\usepackage{url}
\usepackage{verbatim}
\usepackage{graphicx}
\usepackage{cite}
\usepackage{xcolor}  
\usepackage[caption=false,font=normalsize,labelfont=sf,textfont=sf]{subfig}

\usepackage{silence}
\begin{document}

\title{Doubling the Number of Connected Devices in Narrow-band Internet of Things while Maintaining System Performance: An STC-based Approach}

\graphicspath{{./pictures/}}

\author{{Abdulwahid Mohammed},
{Mohammed. S. El-Bakry}, {Hassan Mostafa}, (\textit{Senior Member, IEEE}), and  
	{Abd El-Hady. A. Ammar}
}



\maketitle
\begin{abstract}
Narrow-band Internet of Things (NB-IoT) is a low-power wide-area network (LPWAN) method that was first launched by the 3rd generation partnership project (3GPP) Rel-13 with the purpose of enabling low-cost, low-power and wide-area cellular connection for the Internet of Things (IoT). As the demand for over-the-air services grows and with the number of linked wireless devices reaching 100 billion, wireless spectrum is becoming scarce, necessitating creative techniques that can increase the number of connected devices within a restricted spectral resource in order to satisfy service needs.  Consequently, it is vital that academics develop efficient solutions to fulfill the quality of service (QoS) criteria of the NB-IoT in the context of 5th generation (5G) and beyond. This study paves the way for 5G networks and beyond to have increased capacity and data rate for NB-IoT.  Whereas, this article suggests a method for increasing the number of connected devices by using a technique known as symbol time compression (STC). The suggested method compresses the occupied bandwidth of each device without increasing complexity, losing data throughput or bit error rate (BER) performance. The STC approach is proposed in the literature to work with the conventional orthogonal frequency-division multiplexing (OFDM) to reduce bandwidth usage by 50\% and improve the peak-to-average power ratio (PAPR). Specifically, An STC-based method is proposed that exploits the unused bandwidth to double the number of connected devices while keeping system performance and complexity. Furthermore, the $\mu$-law companding technique is leveraged to reduce the PAPR of the transmitted signals. The obtained simulation results reveal that the proposed approach using the $\mu$-law companding technique increases the transmitted data by twice and reduces the PAPR by 3.22 dB while maintaining the same complexity and bit error rate (BER).  
\end{abstract}


\section{\textbf{Introduction}}
\begin{figure}[!ht]
 	\centering
 	\includegraphics[width=\linewidth,height= 5.2 cm]{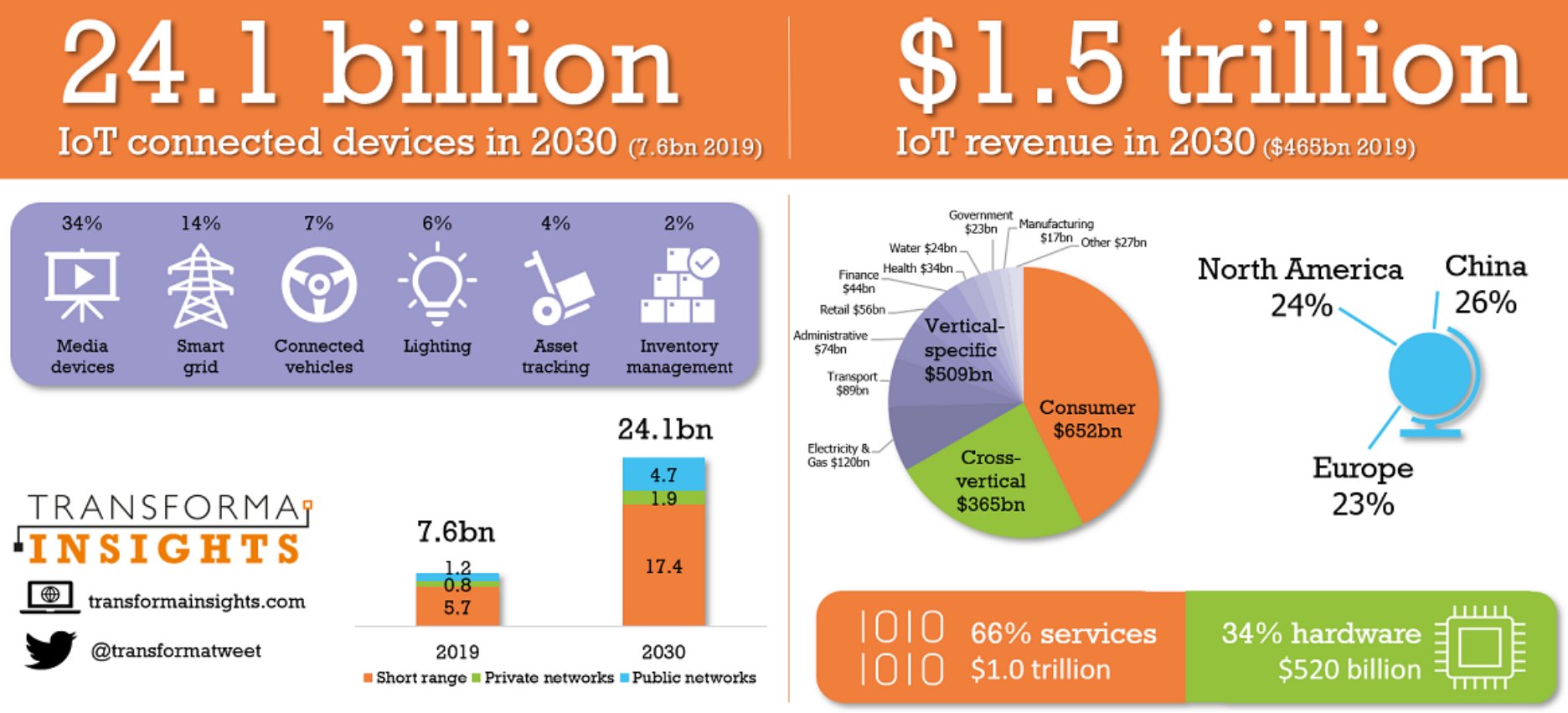}
 	\caption{\label{Infographic} The Internet of Things (IOT) market 2019-2030 (Transforma Insights, 2020) \cite{IoTmarket}}.	
 \end{figure}

\IEEEPARstart{T}{he} Internet of Things (IoT) has changed dramatically in recent years. In particular, the number of IoT devices is rapidly expanding, and various novel IoT applications relating to automobiles, transportation, power grid, agriculture, metering, and other areas have emerged. According to the Forbes analysis 2017, the worldwide IoT industry will be worth \$457 billion by 2020 \cite{IoForecasts}. However, this amount was already surpassed in 2019 with worldwide market size of \$465 billion (Transforma Insights, 2020) \cite{IoTmarket}. According to the recent study provided by \cite{IoTmarket}, there were 7.6 billion active IoT devices at the end of 2019, which is expected to rise to 24.1 billion by 2030, representing an 11\% compound annual growth rate (CAGR), as shown in Fig. \ref{Infographic}. 
 
\begin{figure}[!ht]
 	\centering
 	\includegraphics[width=1\linewidth,height= 3. cm]{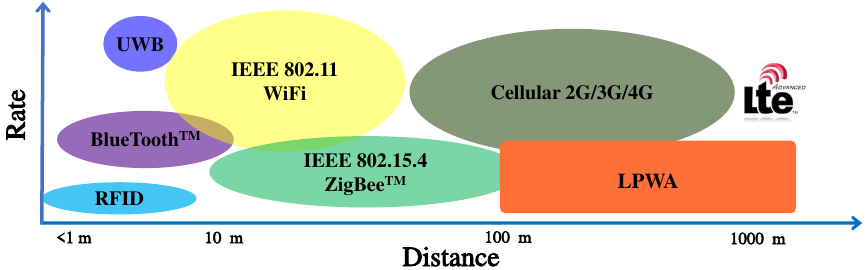}
 	\caption{\label{Coverage} The Coverage and transmission rate comparisons of wireless communication systems \cite{chen2017narrowband}.}
 	\end{figure}

 To meet this enormous need for data, several Low Power Wide Area (LPWA) technologies have been developed with the aim of expanding network coverage, improving power consumption, supporting more users, and reducing device complexity are all supported by these technologies. Different standard development organizations, including  IEEE and 3GPP, work to standardize LPWA technologies. 
 Both, Cellular or non-cellular wireless technology can be used in LPWA. Machine Type Communication (MTC), improved Machine Type Communication (eMTC) and Narrowband Internet of Things (NB-IoT) are examples of cellular technologies, while non-cellular technologies include Long Range (LoRa), ZigBee, Bluetooth, Z-Wave, and others\cite{rastogi2020narrowband}. With the fast development of 5G new radio technologies, intensive research on enhanced mobile broadband (eMBB), massive machine-type communications (mMTCs), and ultrareliable low latency communications (URLLCs) has gotten a lot of interest from academics and industry \cite{liu2019eliminating}. In order to satisfy the 5G outlook, it is necessary not only to achieve significant improvements in new wireless technologies, but also to take into account the harmonious and fair coexistence of heterogeneous networks and  compatibility between 4G and 5G systems\cite{niu2015exploiting}. As shown in Fig. \ref{Coverage}, the unprecedented growth of IoT creates enormous demand for machine-type communications (MTC), which can be divided into three categories: 1) short-distance MTC (distance = 10 m), 2) medium-distance MTC (distance ranges between [10 m, 100 m]), and 3) long-distance MTC (distance $\geq$ 100 m).
 
The remainder of the paper is structured as follows.  Section II  presents the literature review and contribution. Section III overviews the NB-IoT. Section IV offers a description of the general mathematical model of the suggested method. The computational complexity of the different approaches and Algorithms are illustrated in Section V. Section VI covers simulation results and discussion for the suggested method. Section VII concludes the paper.

\section{\textbf{Literature Review and Contribution}}
 The maximum likelihood cross-correlation detection is presented by the authors in \cite{kroll2017maximum} as a hardware implementation. The detector attains an average detection delay that can minimize the power needed per time acquisition by up to 34\%. However, it is a computationally difficult detection approach. The authors of \cite{liu2017investigation} provide a technique for NB-IoT UE to decrease  power consumption during paging loading and offloading. The suggested approach in \cite{liu2017investigation} has the potential to decrease power consumption by around 80\% and enhance energy utilisation efficiency by about 30.5\%. But in standalone mode, this method is not applicable. In \cite{bello2018energy}, a semi-Markov chain model with four states, namely, power saving mode (PSM), idle, random access (RACH), and transmission (Tx) states is developed to evaluate the NB-IoT energy consumption and delay for periodic up-link traffic. However, the model does not account for the energy used while switching between the four modes listed above or the repetition effect on the power consumption. In \cite{adhikary2016performance}, the authors compare NB-IoT coverage under various conditions using 15 kHz and 3.75 kHz spacing. When compared to present LTE technology, the coverage improvement is greater than 20 dB. However, the reported 170 dB of realized maximum coupling loss (MCL) of NB-IoT does not take into account the impairments of channel estimation, carrier offset, or mobility with regard to various configurations. The authors of \cite{andres2018analytic} propose the Link adaptation algorithm, which uses the Shannon theorem's to improve the coverage through characterizing signal-to-noise ratio (SNR), repetition number, and NB-IoT supported bandwidth. Nonetheless, the influence of channel state information on User Equipment (UE) link adaptability was not considered in this study. In \cite{xu2018non}, the authors propose a method  in order to double the number of connected devices by utilizing Fast-OFDM. In comparison to a standard OFDM system, the Fast-OFDM approach decreases the distance between sub-carriers in half, saves 50\% of the bandwidth, and prevents the BER from degrading. However, the proposed approach would result in a mismatch in sampling rate and a carrier frequency offset (CFO). Furthermore, this method is still plagued by the PAPR issue. Unlike the previous studies,  the suggested method in this work improves the system performance by reducing the PAPR issue.\\

 
In summary, the advantages of our suggested method is itemized as follow:
\begin{itemize}
	\item It can decrease the used bandwidth in half by halving the number of sub-carriers, allowing it to transmit twice as much data as a conventional OFDM system.
   \item When compared to the Fast-OFDM and conventional OFDM system, it enhances system performance by minimizing the PAPR issue.
   \item It preserves system performance by preventing BER from deteriorating, as the BER of our suggested technique is precisely equivalent to the BER in the Fast-OFDM and conventional OFDM system.
   \item Although it can transport twice as much data as a standard OFDM system, it has the same complexity.
   \item Since it does not decrease the space between the sub-carriers, it does not result in a sampling rate mismatch or carrier frequency offset (CFO) as in the Fast-OFDM.

\end{itemize}


\section{\textbf{Overview of NB-IoT}}
\begin{figure}[!ht]
	\centering
	\includegraphics[width=0.75\linewidth,height= 12. cm]{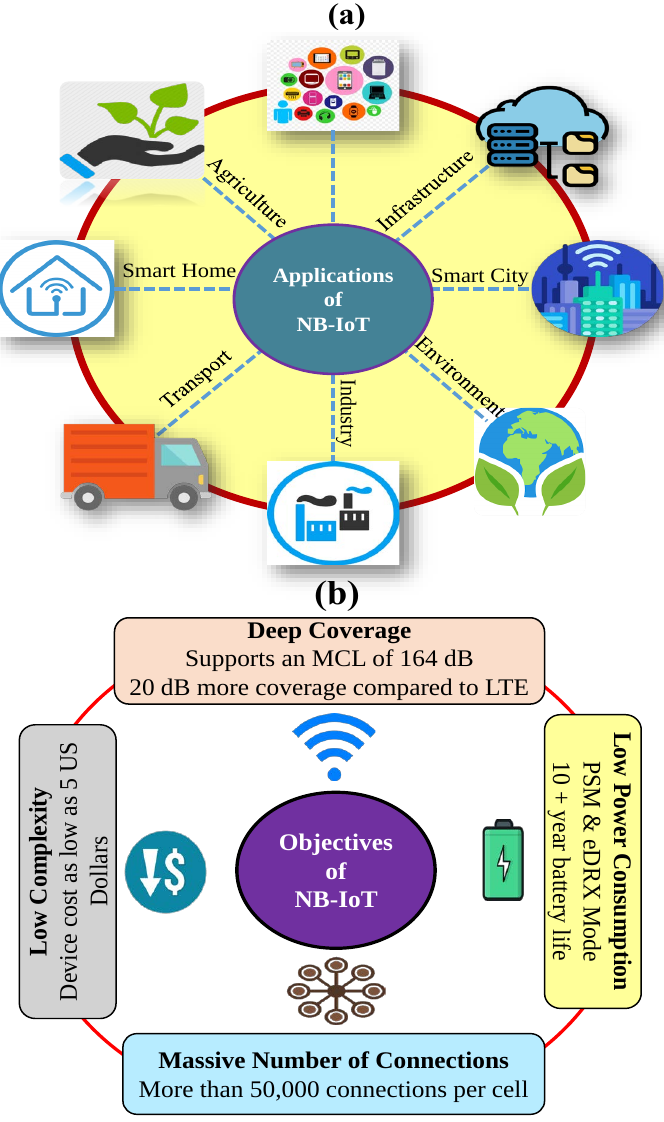}
	\caption{\label{Apps_objs} Applications and Objectives for NB-IOT: (a) Applications for NB-IoT and (b) Objectives for NB-IOT.}	
\end{figure}

For next-generation use cases and applications, the  NB-IoT provides LPWA coverage via massive devices\cite{tusha2018iqi}. NB-IoT is expected to be one of the technologies of 5G new radio (NR) networks, according to\cite{akpakwu2017survey}. Fig. \ref{Apps_objs}(a) shows the applications of NB-IoT such as: smart buildings \cite{qolomany2019leveraging}, smart cities \cite{javidroozi2019urban}, intelligent or smart environmental monitoring system \cite{du2018sensable}, smart metering \cite{wan2019demonstrability}, and intelligent user services \cite{nair2019optimisation}. Also, Smart house, smart wearable gadgets, smart people tracking, and other intelligent and smart user services. Pollution monitoring, intelligent agriculture, water quality monitoring, soil detection, and other aspects of the intelligent or smart environment monitoring system are described in \cite{zhong2016software,wang2016mobile}. The main goals of NB-IoT are outlined in 3GPP specifications \cite{rastogi2020narrowband} and depicted in Fig \ref{Apps_objs}(b).
\begin{itemize}
	{\color{purple}\item \textbf{Deep Coverage:}} The NB-IoT technology is designed to have both indoor and outdoor deep coverage. When compared to the conventional LTE network, NB-IoT provides up to 20 dB higher coverage \cite{beyene2017nb}.  NB-IoT has a maximum coupling loss (MCL) of 164 dB, whereas standard LTE has an MCL of 144 dB. High coverage may be obtained, according to 3GPP specifications, by reducing bandwidth and increasing the number of data transmission repeats. Reduced bandwidth improves the PSD of the user, resulting in increased coverage. Nevertheless, there are two drawbacks to expanding coverage. Reduced bandwidth affects data throughput, whereas a high number of repeats raises data transfer delay and energy consumption \cite{xu2017narrowband}.
	
	{\color{purple}\item \textbf{Low Power Consumption:}} NB-IoT products are designed to have a battery life of much more than ten years. To guarantee that NB-IoT products have such a better battery life, 3GPP Rel-12 and Rel-13 developed Power Saving Mode (PSM) and improved Discontinuous Reception (eDRX) modes. Both of these techniques strive to improve battery life in NB-IoT products by putting them into sleep mode when no data transfer is needed. PSM saves a significant amount of power and has a total sleep duration of 310 hours \cite{sharma2019toward}.
	
	{\color{purple}\item \textbf{Low Complexity:}} The cost of an NB-IoT device must be maintained under \$5 USD. The  infrastructure of  NB-IoT has been simplified and improved in order to decrease devices price. Compared to the conventional LTE system, the network protocol volume, also the number of channels, signals, and transceivers, are reduced in NB-IoT. For both up-link and down-link transmissions, just one transceiver is used \cite{beyene2017performance}. As NB-IoT only supports low-data-rate applications, it does not need a large memory, this leads to decrease the cost of devices.
	
	{\color{purple}\item \textbf{Support of Massive Number of Connections:}} The goal of NB-IoT is to serve over 50,000 users per cell. Using NB-IOT system, the users transfer only low data rate and delay-tolerant data. As a result, a single cell may effectively serve to a massive number of devices. Furthermore, NB-IoT up-link transmission employs sub-carrier level transmission, which improves up-link resource use. For single-tone transmission, NB-IoT offers two numerologies: 15 and 3.75 kHz. When compared to conventional LTE, the signalling overhead of NB-IoT is also simplified \cite{marini2022low}.
\end{itemize}

\section{\textbf{System Model of the Suggested Method}}
\begin{figure*}[!ht]
	\centering
	\includegraphics[width=\linewidth,height= 7.0 cm]{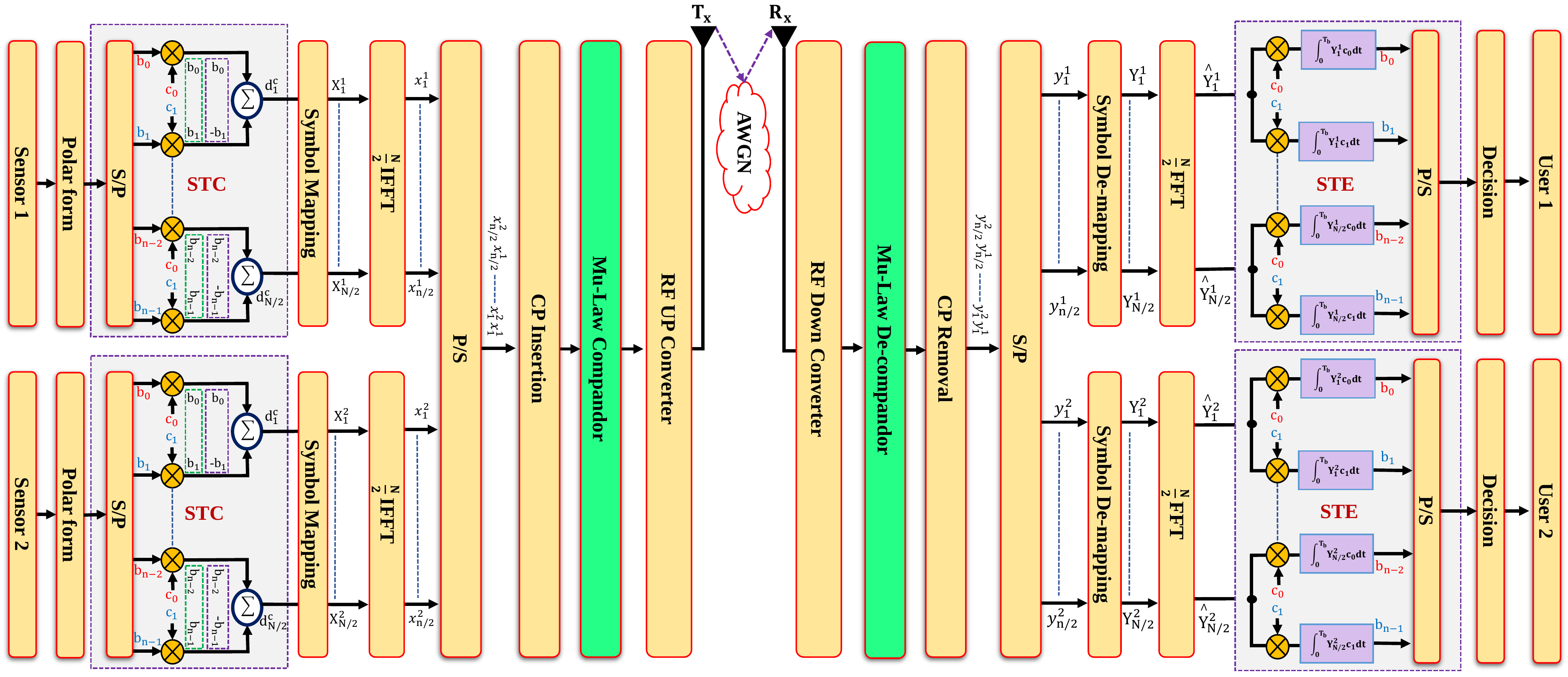}
	\caption{\label{System_Model_TxRx} The transceiver system model for the proposed method using STC and $\mu-$law companding technique.}	
\end{figure*}

Fig. \ref{System_Model_TxRx} shows the transceiver system model for the proposed method using the STC technique with a typical OFDM system (STC-OFDM). The STC-OFDM was initially presented as a wireless approach in \cite{el2017time}, which was proven to have similar results to OFDM while saving 50\% of bandwidth by compressing OFDM symbols by half. In particular, the STC technique uses a spreading and combining mechanism in the transmitter and a symbol time extension (STE) approach in the receiver to expand the received symbol\cite{elbakry2022throughput}, as shown in Fig.\ref{System_Model_TxRx}. First, the transmitted bits are converted into polar forms $b_0$ and $b_1$. Then, using Walsh codes $c_0$ and $c_1$, these polar forms are spread out. Finally, the spread data ,$b_0c_0$ and $b_1c_1$, are combined. The spreading procedure is accomplished by the use of two spreading codes, which are as follows \cite{el2017symbol}:
\begin{equation}
c_0 =[1 \quad  1]   \quad and \quad c_1=[1 \quad -1].
\end{equation}

Table \ref{table1} shows the spreading and combining operations in the STC scheme's transmitter. 
\begin{table}[!ht]
	\centering
	\caption{Spreading and combining process}
	\includegraphics[width=0.85\linewidth,height=1.6 cm]{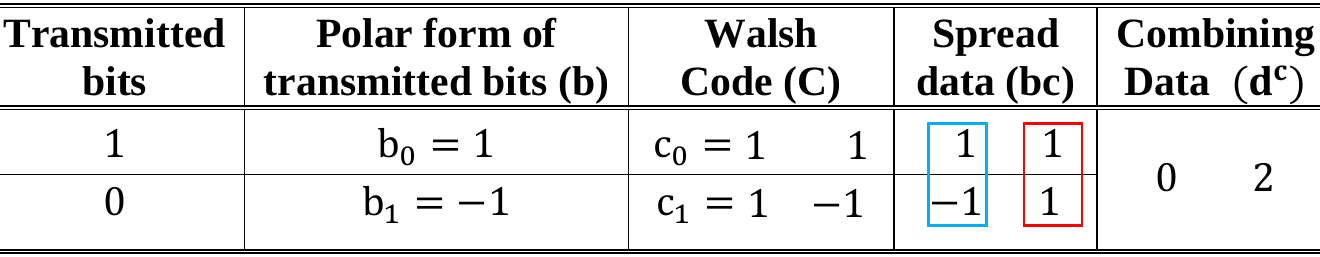}
	\label{table1}
\end{table}

Following the spreading procedure, the spread data streams for each couple are joined as follows:
\begin{equation}
\begin{aligned}
\mathrm{d}_{1}^{\mathrm{c}} &=\mathrm{b}_{1} \mathrm{c}_{0}+\mathrm{b}_{2} \mathrm{c}_{1} \\
\mathrm{~d}_{2}^{\mathrm{c}} &=\mathrm{b}_{3} \mathrm{c}_{0}+\mathrm{b}_{4} \mathrm{c}_{1} \\
& \cdot \\
\mathrm{d}_{\mathrm{N} / 2}^{\mathrm{c}}&=\mathrm{b}_{\mathrm{N}-1} \mathrm{c}_{0}+\mathrm{b}_{\mathrm{N}} \mathrm{c}_{1}.
\end{aligned}
\end{equation}

\begin{figure}[!ht]
	\centering
	\includegraphics[width=0.99\linewidth,height= 4.0 cm]{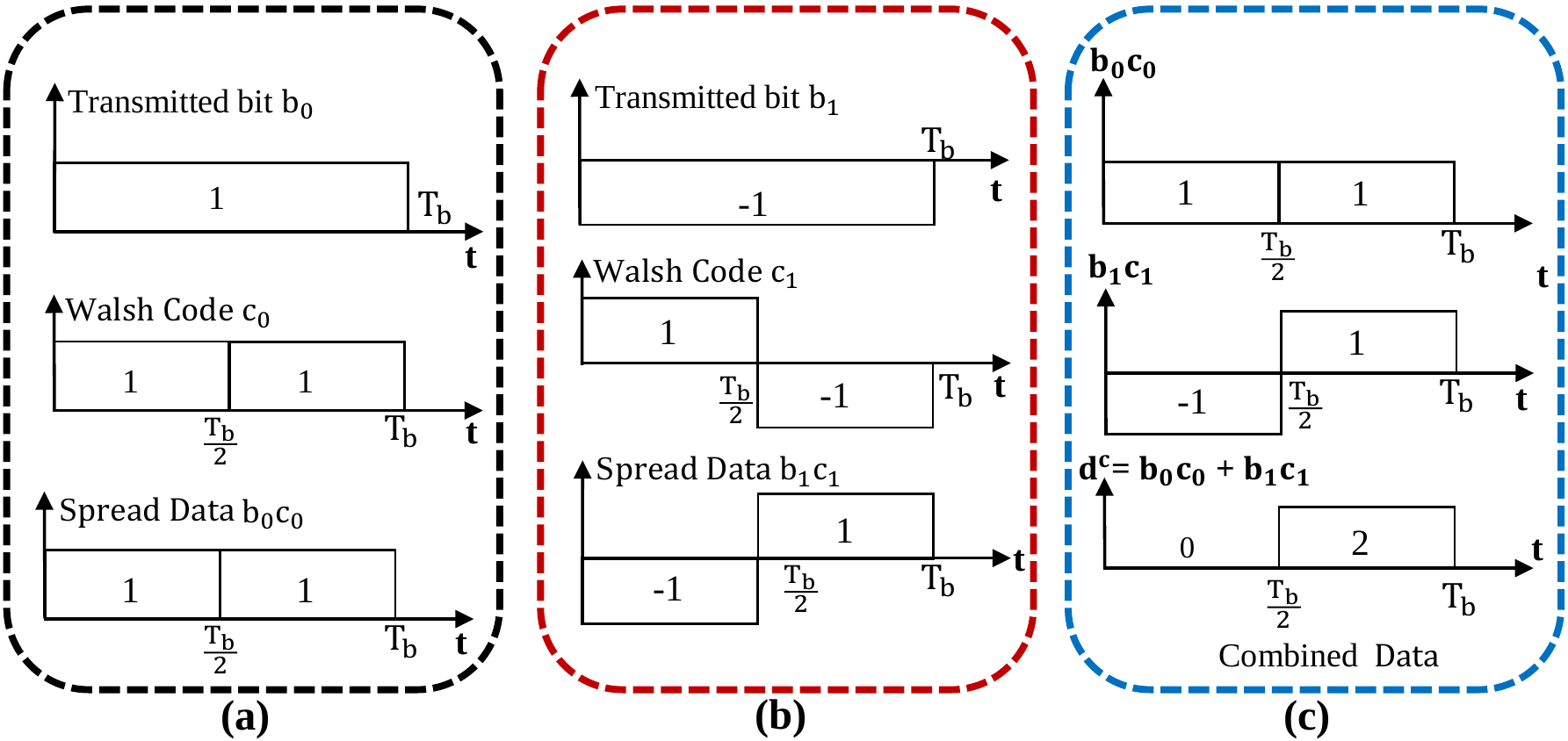}
	\caption{\label{Waveforms_of_transmitted_data} The wave-forms for: (a) spreading process of the transmitted data $b_0$, (b) spreading process of the transmitted data $b_1$ and (c) combined data}	
\end{figure}

Where $b$ represents a bit stream, $c$ is Walsh spreading code with chip rate $Tc = Tb/2$ and $d^c$ denotes the combining  data symbol. Fig. \ref{Waveforms_of_transmitted_data} (a) depicts the waveform of spreading process ($b_0c_0$) for the transmitted data $b_0$ and Fig. \ref{Waveforms_of_transmitted_data} (b) displays the waveform of spreading process ($b_1c_1$) for the transmitted data $b_1$. While, Fig. \ref{Waveforms_of_transmitted_data} (c) shows the combined data  ($d^c$). As displayed in Fig. \ref{System_Model_TxRx}, the combining symbols are passed through the symbol mapping block to get the complex data symbols. The complex data symbol on the $K^{th}$ sub-carrier is denoted by $X_k^c$, $k=1,2,..., N/2$. The resulting $N/2$ signals are applied to the $N/2$ input ports of an inverse fast Fourier transform (IFFT) processor. A discrete-time OFDM symbol after IFFT is expressed as follows \cite{el2017symbol}:
\begin{equation}
x_{k}^{c}=\frac{2}{N} \sum_{m=0}^{\frac{N}{2}-1} X_{m}^{c} e^{j 2 \pi k m /\frac{N}{2}}, \quad 0 \leq k \leq \frac{N}{2}-1
\end{equation}

Where $k$ represents the time index, $N$ is the number of sub-carriers, $x_{k}^{c}$ is the $k^{th}$ OFDM symbol and $X_m$ represents the  $m^{th}$ transmitted data symbols. The cyclic prefix (CP) is inserted in front of each OFDM signal as a guard interval (GI) between subsequent OFDM symbols in OFDM systems to keep the orthogonality criterion and prevent inter-carrier interference (ICI) and inter-symbol interference (ISI). Where the length of CP must be greater than the maximum multi-path channel delay spread ($T_{cp} \geq \tau_{max}$). A parallel-to-serial (P/S) converter is applied to the resulting time domain symbols.  A  cyclic prefix (CP) of a suitable length ($L_{cp}$) is added to combat the effect of multi-path propagation and the transmitted OFDM symbol with CP is defined as follows:

\begin{equation}
x_{k}^{c}(cp)=\frac{2}{N} \sum_{m=0}^{\frac{N}{2}-1} X_{m}^{c} e^{j 2 \pi k m /\frac{N}{2}},  -L_{cp} \leq k \leq \frac{N}{2}-1
\end{equation}

To recover the transmitted data, the transmitter processing is functionally reversed in reverse order at the receiver side, as shown in Fig. \ref{System_Model_TxRx}, to yield an approximated form of the binary information sequence. The $k^{th}$ received compressed OFDM symbol in frequency domain  is expressed as follow: 
\begin{equation}
Y_{k}^{c}=\frac{2}{N} \sum_{m=0}^{\frac{N}{2}-1} y_{m}^{c} e^{-j 2 \pi k m /\frac{N}{2}}, \quad 0 \leq k \leq \frac{N}{2}-1
\end{equation}

Following symbol de-mapping, the STE technique is used to retrieve the data streams, as shown in  Table \ref{table2}.
\begin{table}[ht!]
	\centering
	\caption{De-spreading and combining processes}
	\includegraphics[width=0.8\linewidth,height=1.2 cm]{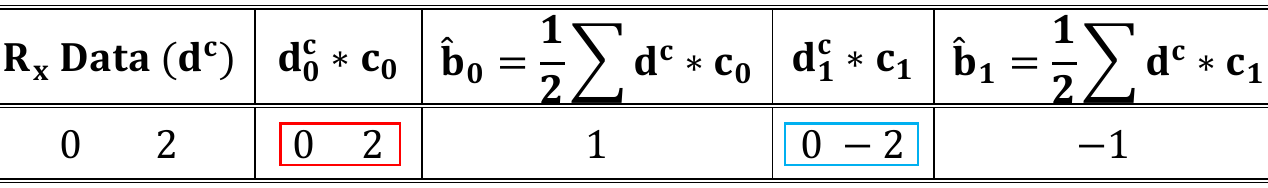}
	\label{table2}
\end{table}

To de-spread the received data, it is multiplied by the same code as the transmitter ($c_0$ and $c_1$). Then, integration throughout the bit period is as follows:
\begin{equation}
\begin{gathered}
\hat{b}_{1}=\int_{0}^{T_{b}} \hat{Y}_{1}^c c_{0} d t, \quad \hat{b}_{2}=\int_{0}^{T_{b}} \hat{Y}_{1}^c c_{1} d t \\
\hat{b}_{3}=\int_{0}^{T_{b}} \hat{Y}_{2}^c c_{0} d t, \quad \hat{b}_{4}=\int_{0}^{T_{b}} \hat{Y}_{2}^c c_{1} d t \\
\vdots \\
\hat{b}_{N-1}=\int_{0}^{T_{b}} \hat{Y}_{N / 2}^c c_{0} d t, \quad \hat{b}_{N}=\int_{0}^{T_{b}} \hat{Y}_{N / 2}^c c_{1} d t
\end{gathered}
\end{equation}
Where $T_b$ denotes the bit duration, $\hat{b}_{N}$ is the detected $N^{th}$ bit and $\hat{Y}_{m}^c$ is the de-mapped symbol.\\

The suggested approach is based on employing the STC technique twice in order to send data in the unused bandwidth, therefore doubling the number of connected devices. On the other hand, STE technique is employed twice on the receiver side to recover the data transferred from the transmitter to its original form, as shown in Fig. \ref{System_Model_TxRx}.

\section{\textbf{Computational Complexity Analysis and Algorithms}}
This section explains the algorithms that utilized for the STC-OFDM and the proposed  method. Algorithm \ref{CST_func} is designed to address the STC scheme and its inverse STC$^{-1}$, while Algorithm \ref{Transceiver_proposed} represents the transceiver  of the suggested method (Transmitter and Receiver). Table \ref{Complex} represents comparison between the computational complexity of the proposed method using $\mu-$law companding technique and the  another techniques. It is clear from Table \ref{Complex} that the complexity of the proposed method using $\mu-$law technique is nearly the same with the conventional OFDM. Consequently, this indicates that the proposed method does not cause an increase in the complexity of the system.

\begin{table}[!ht]
	\renewcommand{\arraystretch}{1.5}
	\centering
	\caption{Computational Complexity Analysis.}
	\resizebox{\columnwidth}{!}{	\begin{tabular}{|c|c|c|}
			\hline
			\textbf{Techniques} & \textbf{ No. Multiplications} & \textbf{ No. Additions} \\
			\hline
			\hline
			Conventional OFDM  \cite{afifi2020efficient}
			& 2N Log$_2$N - 2N 
			& 3N Log$_2$N - N  \\
			\hline
			Fast OFDM  
			& 2N Log$_2$N - 2N 
			& 3N Log$_2$N - N  \\
			\hline
			STC-OFDM  \cite{afifi2020efficient}
			& N Log$_2$$\frac{N}{2}$ - N 
			& $\frac{3}{2}$N Log$_2$$\frac{N}{2}$ - $\frac{N}{2}$   \\
			\hline
			The proposed scheme 
			& 2N Log$_2$N - 2N 
			& 3N Log$_2$N - N  \\
			\hline
			The proposed scheme using  $\mu$-law \cite{mohammed2021novel}
			& 2N Log$_2$N - N 
			& 3N Log$_2$N + 3N  \\
			\hline
	\end{tabular}}
	\label{Complex}
\end{table}

\begin{algorithm}[!ht]
	\caption{: STC \& STC$^{-1}$  functions}
	\label{CST_func}
	\begin{algorithmic}
		\STATE \textbf{At the Transmitter STC function}
	\end{algorithmic}
	\begin{algorithmic}[1]
		\STATE D$= reshape(Data, length(D)/2,2)$; $\rightarrow$ Converting generated data from serial to parallel
		(Polar Matrix form) 
		\STATE W$ = hadamard(n)$;$\rightarrow$ Generate walsh code \& n = 2
		\STATE SpreadData $= D*W$; $\rightarrow$ Spreading Data
		\STATE CombData$ = comb(SpreadData)$; $\rightarrow$ Combining Data
		\STATE Norm$ = CombineData/2$; 		
		\STATE $f(X) =$  Norm(:,1) + j*Norm(:,2); 					
		
	\end{algorithmic}
	\begin{algorithmic}
		\STATE \textbf{At the Receiver STC$^{-1}$  function}
	\end{algorithmic}
	\begin{algorithmic}[1]
		\STATE W$ = hadamard(n)$;$\rightarrow$ Generate walsh code \& n = 2
		\STATE R$_x$ = Received Data in complex form ; 
		\STATE RealR$_x$ = real (R$_x$); $\rightarrow$ Peal Part 
		\STATE ImagR$_x$ = imag (R$_x$); $\rightarrow$ Imaginary part 
		\STATE R$_y$ =  $ [\;RealR_x ; ImagR_x\;]$; 
		\STATE Despread$_1$ = R$_y$*W(1,:); $\rightarrow$ De-spreading Data  		 
		\STATE Despread$_2$ = R$_y$*W(2,:); $\rightarrow$ De-spreading Data  	
		\FOR{$k = 1:length (R_y)$}	
		\STATE CombData$_1$ = sum (Despread$_1$(k,:)) $\rightarrow$ Combining Data  		 
		\STATE CombData$_2$ = sum (Despread$_2$(k,:)) $\rightarrow$ Combining Data  	
		\ENDFOR		
		\STATE Y =  $ [\;CombData_1 ; CombData_2\;]$; 
		
		\STATE $f^{-1}(X) = $ (Y+1)/2; $\rightarrow$ 
	\end{algorithmic}
\end{algorithm}

\begin{algorithm}[!ht]
	\caption{: The Transceiver of Proposed Technique}
		\label{Transceiver_proposed}
	\begin{algorithmic}[1]
		\STATE Error = zeros $(1,length(EbN0dB))$;
		\STATE nloop = 100;	$\rightarrow$ Number of simulation loops 	
		\STATE nsym = 1000;	$\rightarrow$ Number of OFDM symbols for one loop
		\STATE EbN0dB = 0:12;	$\rightarrow$ Bit to noise ratio (Eb/N0)	 
		\FOR{$i = 1:nloop$}	
		\STATE D$_1$ = Generate data for first source  
		\STATE D$_2$ = Generate data for second source  
		\STATE X$_1$ = STC(D$_1$) $\rightarrow$ Apply STC Tech on the 1$^{st}$ source 
		\STATE X$_2$ = STC(D$_1$) $\rightarrow$ Apply STC Tech on the 2$^{nd}$ source 		 
		\STATE $x_1$ = \textbf{IFFT}$(X_1)$; $\rightarrow$ Convert to time domain
		\STATE $x_2$ = \textbf{IFFT}$(X_2)$; $\rightarrow$ Convert to time domain
		\STATE $y_1$ = $x_1 +$ $cp_1$;  $\rightarrow$ Add cyclic prefix 
		\STATE $y_2$ = $x_2 +$ $cp_2$;  $\rightarrow$ Add cyclic prefix 
		\STATE $y$ =  $ [\;y_1 ; y_2\;]$;  $\rightarrow$ Data of the two sources 	
		\STATE PAPR = zeros $(1,N)$; $\rightarrow$ $N$ is the number of IFFT,		
		\FOR{$k = 1:nsym$}	
		\STATE Peak-power = \textbf{max}$(|y|^2)$; 
		\STATE Avg-power = \textbf{mean}$(|y|^2)$;
		\STATE PAPR($k$) = $10\times$log$_{10}(\text{Peak-power}/\text{Avg-power})$;
		\ENDFOR
		\FOR{$q = 1:length(EsN0dB)$}	
		\STATE $r_x = y +$ noise;  $\rightarrow$ Received under AWGN channel 
		\STATE $r_{cp1}=$ RemoveCP ($r_{x1}$);  $\rightarrow$Remove CP of the 1$^{st}$ source
		\STATE $r_{cp2}=  $RemoveCP ($r_{x2}$);  $\rightarrow$Remove CP of the 2$^{nd}$ source
		\STATE $R_{y1} = $\textbf{FFT}$(r_{cp1})$; $\rightarrow$     Convert to frequency domain
		\STATE $R_{y2} = $\textbf{FFT}$(r_{cp2})$; $\rightarrow$     Convert to frequency domain
		\STATE $R_{y1} = $\textbf{STC}$^{-1}(r_{cp1})$; $\rightarrow$ Apply the inverse STC Tech to 1$^{st}$ source
		\STATE $R_{y2} = $\textbf{STC}$^{-1}(r_{cp2})$; $\rightarrow$Apply the inverse STC Tech to 2$^{nd}$ source
		\STATE $R_y$ =  $ [\;R_{y1} ; R_{y2}\;]$;  $\rightarrow$ Data of the two sources 	
		\FOR{$m = 1:length(R_y)$}	
		\IF {$R_y(m) > 0.5$}
		\STATE $R_y(m) = 1$
		\ELSE
		\STATE  $R_y(m) = 0$
		\ENDIF          
		\ENDFOR	
		\STATE Error $=$ BER ($y$ , Output); $\rightarrow$ BER calculation
		\ENDFOR
		\STATE D$_1$ = R $_{y}(1:length(R_y/2))$;  $\rightarrow$ Received Data of the 1$^{st}$ source
		\STATE D$_2$ = R$_{y}$($length(R_y/2)+1 : end$);  $\rightarrow$ Received Data of the 2$^{nd}$ source
		\ENDFOR
	\end{algorithmic}
\end{algorithm}

\section{\textbf{Simulation Results and Discussion}}
This section discusses the numerical simulation and results for the proposed design. Table (\ref{parametter}) lists the used simulation parameters of the system. Improvement in PAPR, degradation in bit error rate (BER), and power spectral density (PSD) are used as performance metrics of interest. Binary Phase Shift Keying (BPSK) modulation is used in this study for Fast-OFDM, STC-OFDM and typical OFDM. Fast-OFDM is only applicable with one-dimensional modulation schemes, as was mentioned in \cite{xu2018non}. Fast-OFDM and STC-OFDM systems are unable to handle higher order modulation forms like Quadrature Phase Shift Keying (QPSK), which NB-IoT supports. Nevertheless, a developed non-orthogonal wave-forms known as "SEFDM" \cite{7572173}, may employ modulation schemes up to 16QAM at the cost of larger and more complexity. The improvement in PAPR is the difference between the PAPR of the original signal (conventional OFDM signal) and the PAPR of the STC-OFDM signal (i.e. Improvement in PAPR = PAPR of original signal - PAPR of STC-OFDM signal). Similarly, the degradation in BER is the difference between the BER of the original signal and the BER of the STC-OFDM signal.
\renewcommand{\arraystretch}{1.}
\small\addtolength{\tabcolsep}{-2. pt}
\begin{table}[!ht]
	\centering
	\caption{Simulation Parameters \cite{xu2018non}}
	\begin{tabular}{l|l|l|l}
		\hline
		\textbf{Parameter }
		&  \textbf{OFDM} 
		& \textbf{Fast-OFDM}
		& \textbf{STC-OFDM}\\
		\hline
		\hline
		
		Occupied Channel BW
		&180 kHz &90 kHz &90 kHz \\		
		Bit rate (kbit/s)
		&180 Kbps &180 Kbps &180 Kbps \\
		Spacing frequency, $\varDelta$f
		&15 kHz &7.5 kHz &15 kHz \\
		Sampling frequency , $f_s$
		& 1.92 MHz & 1.92 MHz & 1.92 MHz \\
		FFT size, $N$
		&128 &128 & 64 \\
		CP length,$N_{CP}$
		&32 samples &32 samples & 16 samples \\
		Modulation type
		& BPSK& BPSK &BPSK \\
		Channel model
		& AWGN & AWGN  & AWGN \\
		\hline
	\end{tabular}%
	\label{parametter}%
\end{table} 

\begin{figure}[!ht]
	\centering
	\includegraphics[width= 1.00\linewidth,height=8.7 cm]{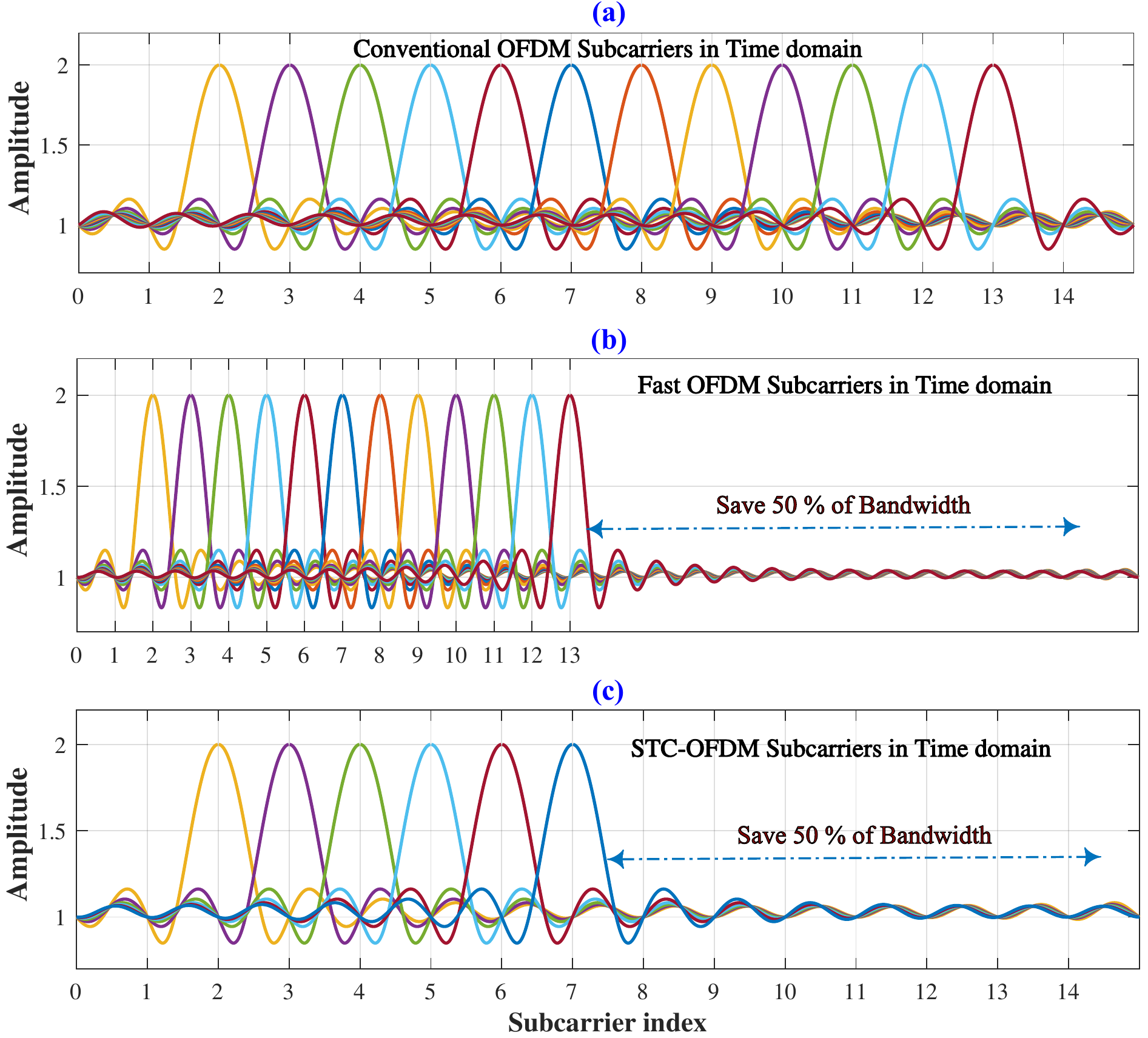}
	\caption{Sub-carrier allocation schemes for: (a) Typical OFDM, (b) Fast OFDM \cite{xu2018non} and (c) STC-OFDM System \cite{el2017symbol}.}
	\label{Subcarrier_OFDM}
\end{figure}

In Fig.\ref{Subcarrier_OFDM}, the sub-carriers for the traditional OFDM technique, Fast-OFDM and the STC-OFDM are compared. Fig. \ref{Subcarrier_OFDM}(a) shows a standard OFDM spectrum with orthogonally packed sub-carriers. The number of sub-carriers for Fast-OFDM is the same with typical OFDM, but the space between the sub-carriers is decreased to half as compared to typical OFDM as depicted in Fig. \ref{Subcarrier_OFDM}(b). STC-OFDM reduces the number of sub-carriers by half while keeping the sub-carrier spacing, as it is clear in Fig. \ref{Subcarrier_OFDM}(c). As a consequence, the bandwidth in Fast-OFDM and STC-OFDM is reduced by half, and the conserved bandwidth might be utilized to power more devices. 

\begin{figure}[!ht]
	\centering
	\includegraphics[width= 0.95\linewidth,height=8.2 cm]{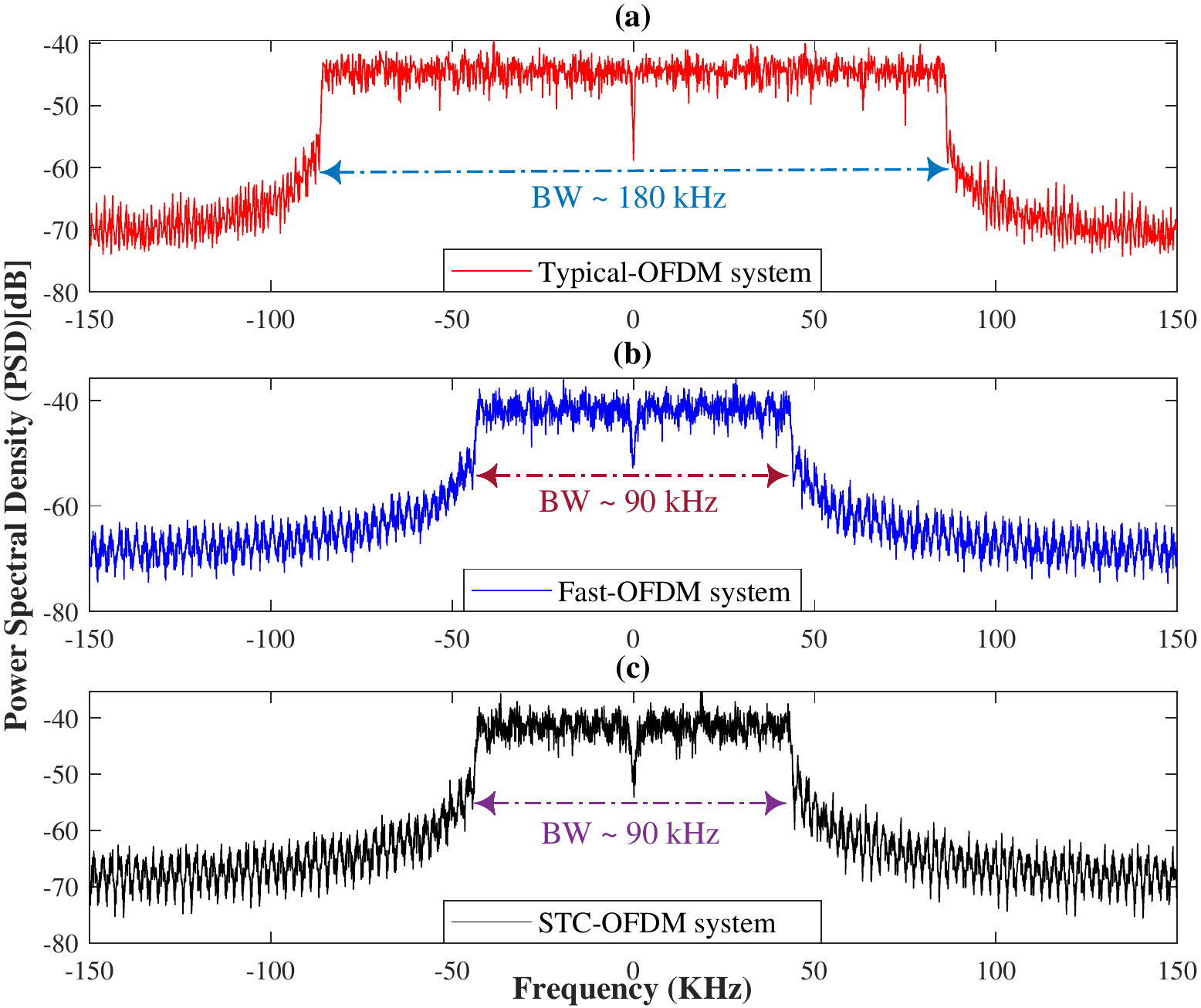}
	\caption{PSD for: (a) Typical OFDM, (b) Fast OFDM \cite{xu2018non} and (c) STC-OFDM System \cite{el2017symbol}.}
	\label{PSD_OFDM_CST}
\end{figure}

Fig. \ref{PSD_OFDM_CST} shows the spectrum of STC-OFDM, Fast-OFDM, and typical OFDM. The first spectrum is for conventional OFDM and offers a bandwidth of 180 kHz; the second is for Fast-OFDM and compresses a bandwidth to 90 kHz; and the third is for STC-OFDM and also has a bandwidth of 90 kHz. 
The Fast-OFDM can transmit the same amount of data with half the bandwidth as compared to typical OFDM by decreasing the sub-carriers spacing to half,  while the STC-OFDM sends the same amount of data with half the bandwidth as compared to typical OFDM by decreasing the number of sub-carriers to half, as it is seen in Fig. \ref{PSD_OFDM_CST}. This means that they can save 50\% of the bandwidth.

\begin{figure}[!ht]
	\centering
	\includegraphics[width=0.9 \linewidth,height=11.2 cm]{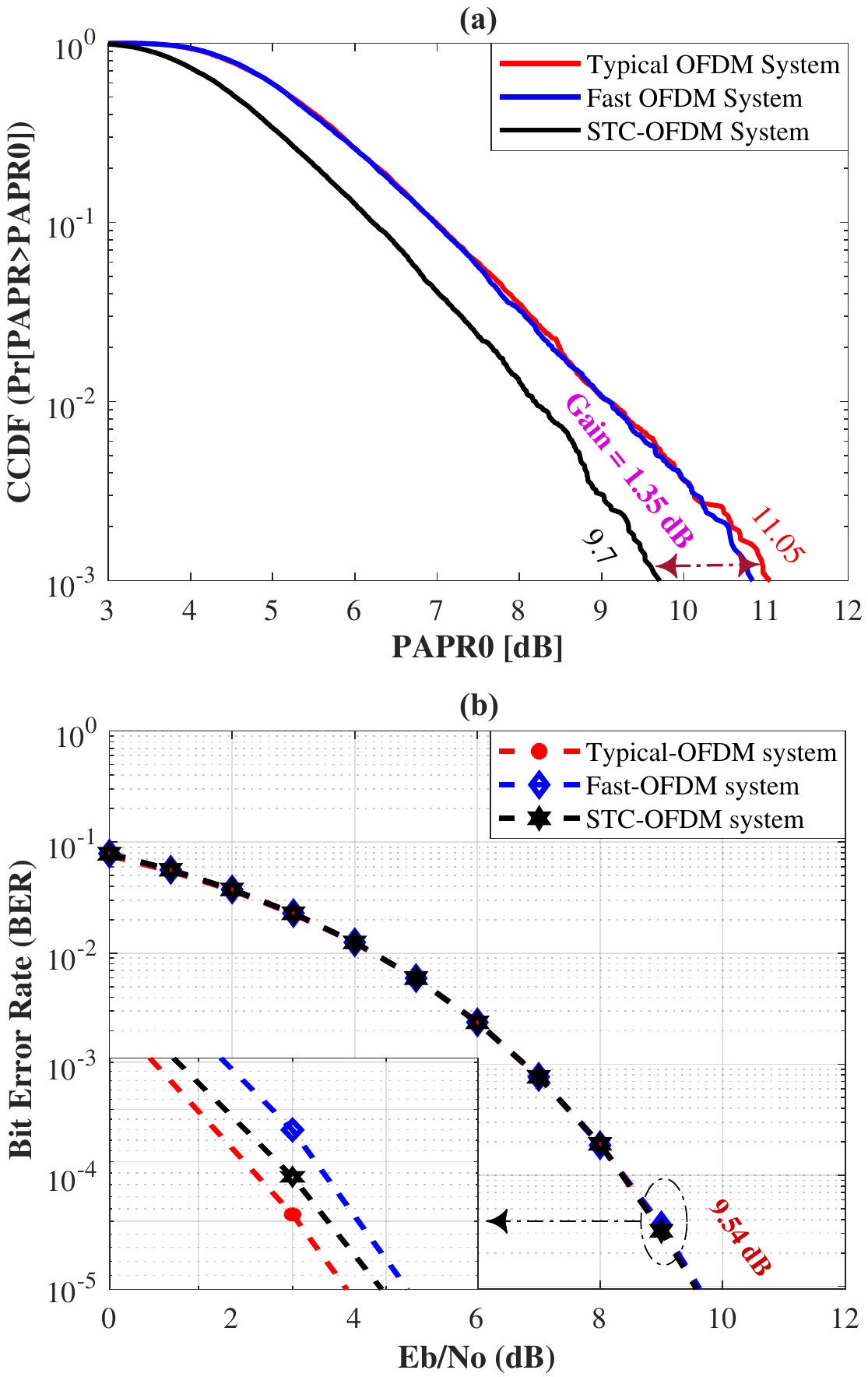}
	\caption{Performance comparison between Typical OFDM, Fast OFDM \cite{xu2018non} and STC-OFDM System \cite{el2017symbol} based on (a) PAPR and (b) BER.}
	\label{PAPR_BER_OFDM_Fast_STC}
\end{figure}
The STC-OFDM system uses half the bandwidth of the typical OFDM system to transmit the same amount of data. As a result, as illustrated in Fig. \ref{PSD_OFDM_CST}(c), 50\% of the bandwidth will be available for use to transmit the same amount of data. Furthermore, the STC-OFDM system increases system performance by decreasing the PAPR issue and avoiding the BER deterioration. Figs. \ref{PAPR_BER_OFDM_Fast_STC}(a) and  \ref{PAPR_BER_OFDM_Fast_STC}(b) indicate that the PAPR gain is 1.35 dB utilizing the STC-OFDM system with no BER deterioration. As for the Fast-OFDM, it  saves half the bandwidth (Fig. \ref{PSD_OFDM_CST}(b)) and  prevents the BER degradation (Fig. \ref{PAPR_BER_OFDM_Fast_STC}(b)), but it does not reduce PAPR, as seen in Fig. \ref{PAPR_BER_OFDM_Fast_STC}(a).\\

\begin{figure}[!ht]
	\centering
	\includegraphics[width=\linewidth,height= 3.2 cm]{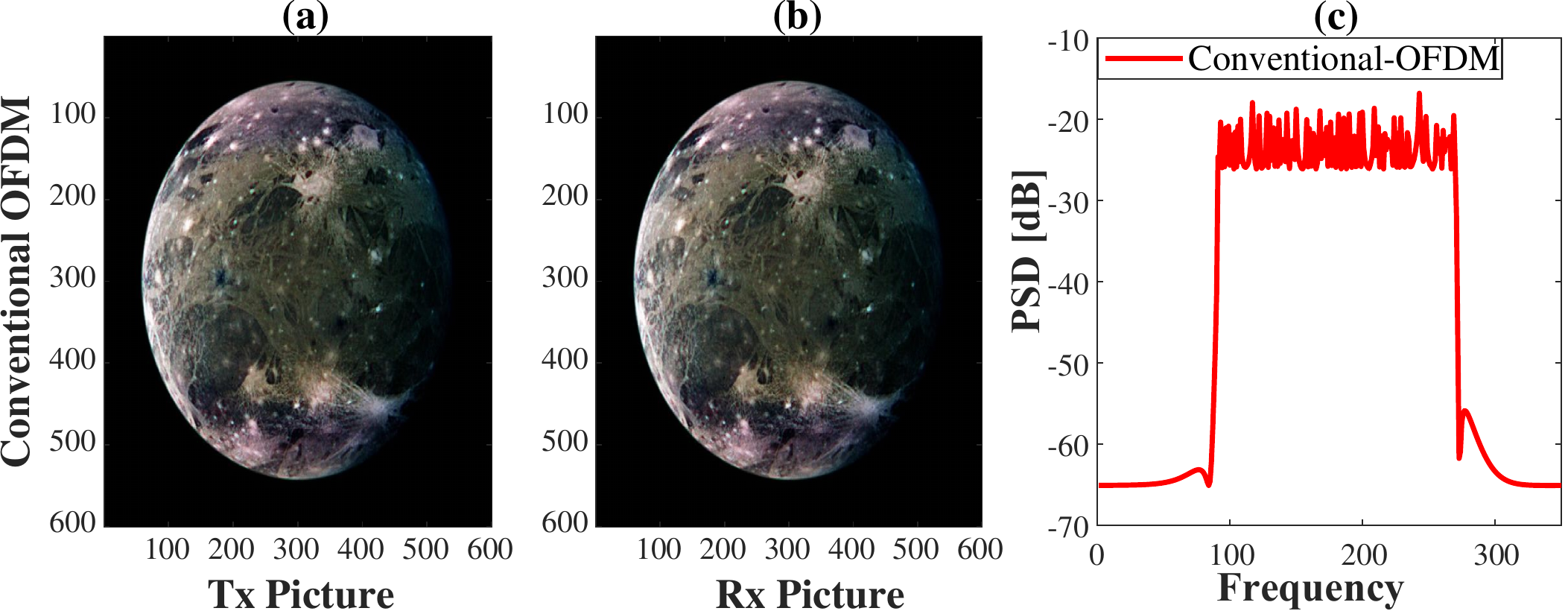}
	\caption{\label{Moon_Pic} Send and receive a picture using conventional OFDM system.}	
\end{figure}

\begin{figure}[!ht]
	\centering
	\includegraphics[width=\linewidth,height= 5. cm]{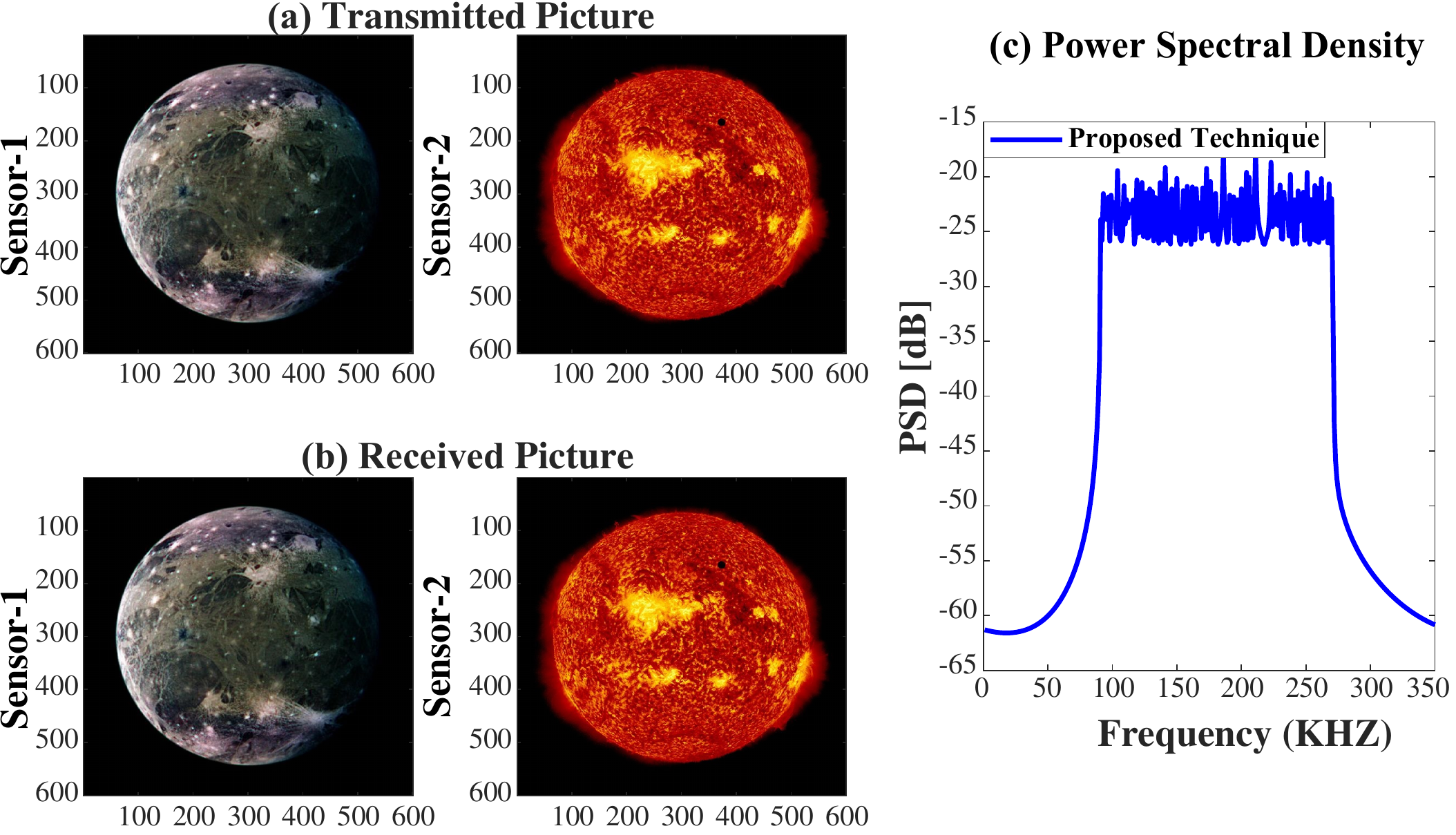}
	\caption{\label{Moon_Sun_Pic} Send and receive two pictures using the proposed technique.}	
\end{figure}

From Fig. \ref{Moon_Pic} and Fig. \ref{Moon_Sun_Pic}, it is clear that the transmitted data can increase to double using the same bandwidth. Fig. \ref{Moon_Pic} shows the transmitted picture, received picture and PSD for conventional OFDM system, while Fig. \ref{Moon_Sun_Pic} displays the transmitted pictures, received pictures and PSD for the proposed scheme. The suggested approach concurrently transmits data from two sources (in our example, two sensors) with the same bandwidth as a standard OFDM system. Fig. \ref{Moon_Pic}(c) and Fig. \ref{Moon_Sun_Pic}(c) illustrate that the traditional OFDM technique and the suggested method have the same bandwidth, but the suggested approach has the benefit of being able to transfer twice as much information as the conventional OFDM technique,  as seen in Figs. \ref{Moon_Pic} and \ref{Moon_Sun_Pic}.

\begin{figure}[!ht]
	\centering
	\includegraphics[width= 0.92\linewidth,height=17.5 cm]{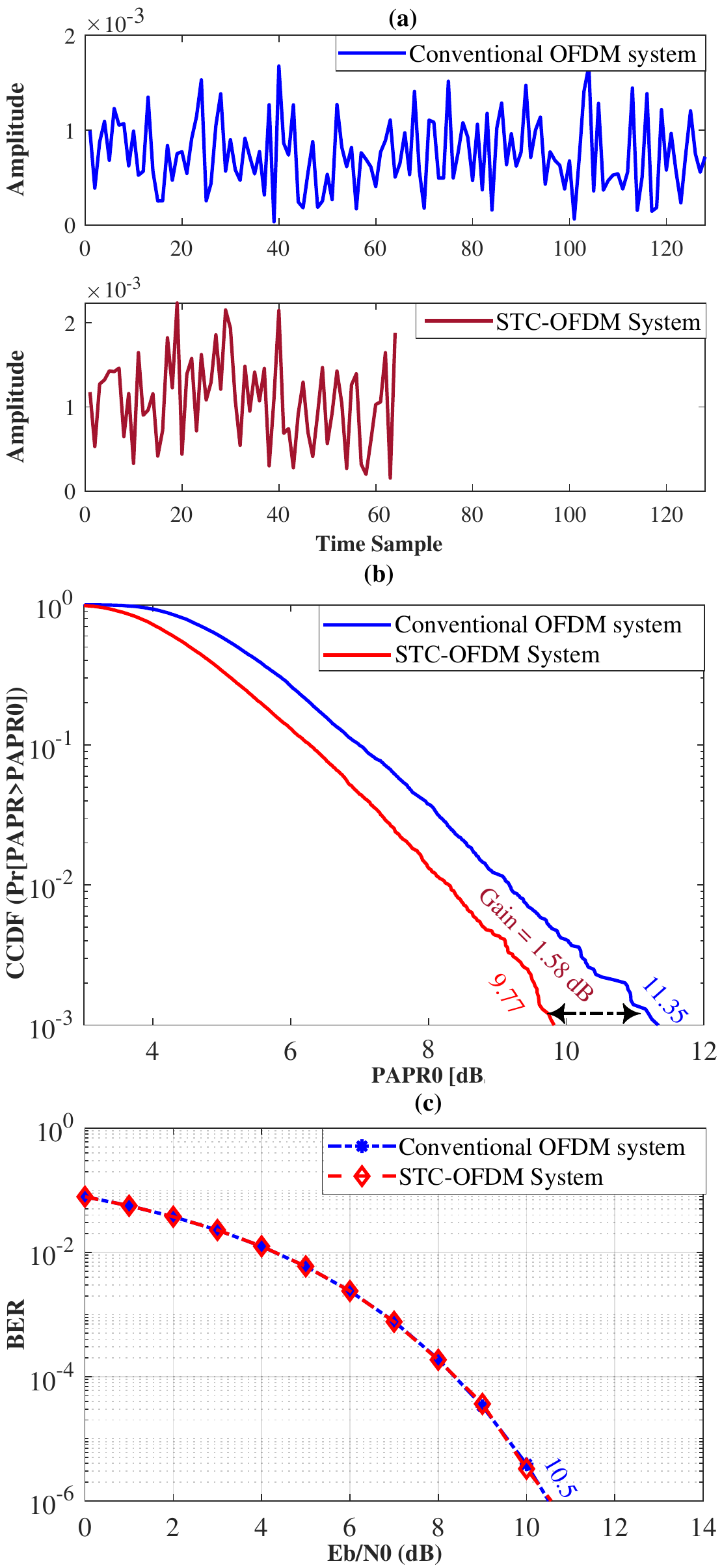}
	\caption{Performance comparison between typical OFDM and STC-OFDM technique \cite{el2017symbol} based on: (a) Time domain of Tx signal, (b) PAPR and (c) BER.}
	\label{Time_PAPR_BER}
\end{figure}

The time domain of the transmitted signal is measured after IFFT in the transmitter side for both the traditional OFDM system and the STC-OFDM method. When compared to a conventional OFDM system, the STC-OFDM scheme reduces OFDM symbol time by half for the same amount of data, as displayed in Fig. \ref{Time_PAPR_BER}(a). The transmitted information of a traditional OFDM system with 128 samples is transferred in 64 samples utilizing the STC-OFDM method, as it is clear in Fig. \ref{Time_PAPR_BER}(a). STC-OFDM is found to have similar performance to conventional OFDM, but with a 50\% bandwidth savings. The number of IFFT is reduced by half when using the STC-OFDM technique and the PAPR is reduced as well. When employing STC-OFDM technique, the PAPR improves by 1.58 dB, where the PAPR of typical OFDM is 11.35 and the PAPR of STC-OFDM is 9.77, as it is depicted in Fig. \ref{Time_PAPR_BER}(b). In addition to the improvement in PAPR, the STC-OFDM approach ensures that the BER of traditional OFDM and STC-OFDM is the same (i.e. the degradation in BER =0), as shown Fig. \ref{Time_PAPR_BER}(c).

\begin{figure}[!ht]
	\centering
	\includegraphics[width= .93\linewidth,height=17. cm]{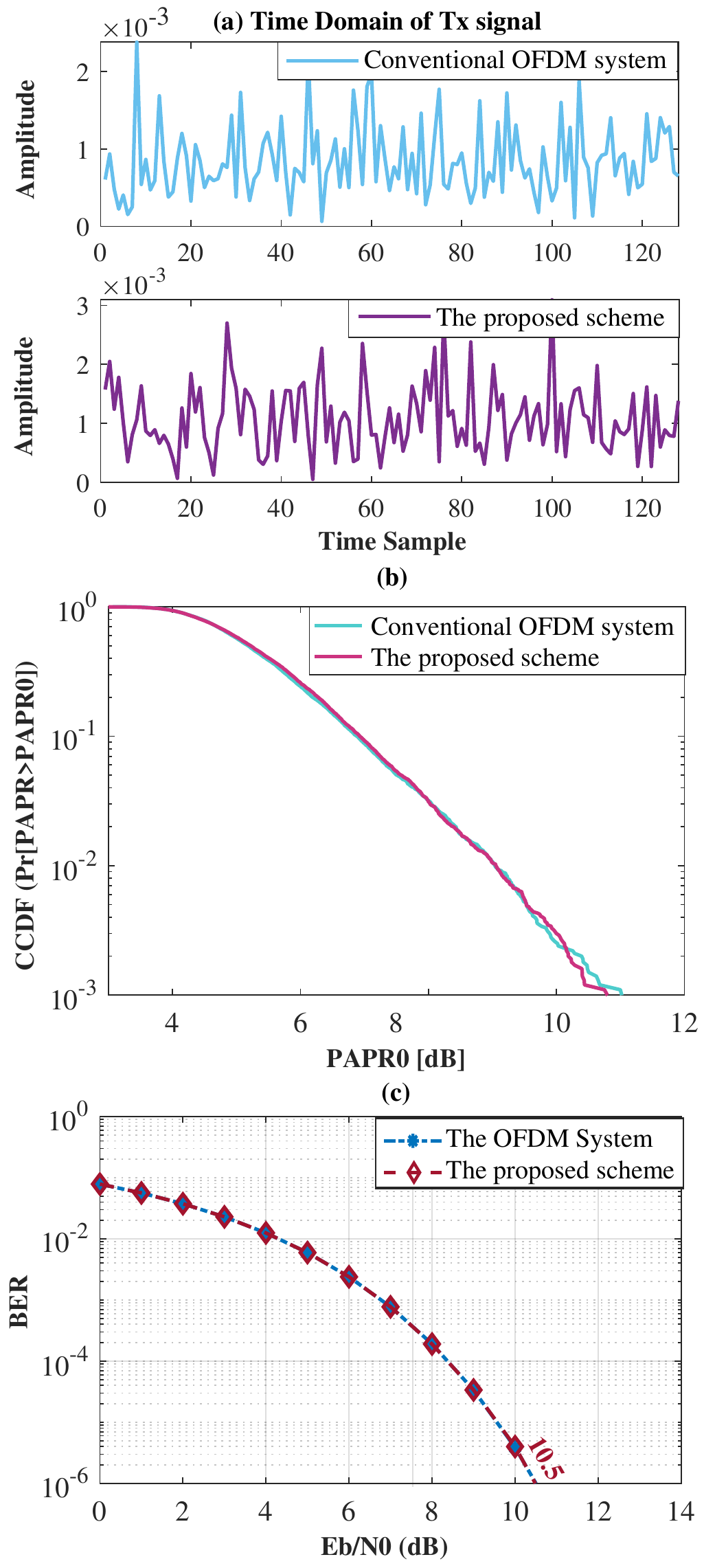}
	\caption{Performance comparison between conventional OFDM and the proposed method based on: (a) Time domain of Tx signal, (b) PAPR and (c) BER.}
	\label{Time_PAPR_BER_2s}
\end{figure}

The STC-OFDM approach can save 50\% of bandwidth, which means there is 50\% of capacity that is not being used. As shown in Fig. \ref{System_Model_TxRx}, the suggested method employs two STC technique to use the entire bandwidth and double the amount of data transferred. Fig. \ref{Time_PAPR_BER_2s}(a) shows that the suggested method has the same OFDM symbol time as a typical OFDM system, but it can send twice as much data, as displayed in Fig. \ref{Moon_Sun_Pic}. The suggested approach offers the same PAPR as traditional OFDM and the same BER performance. The PAPR of a typical OFDM system and the suggested approach are shown in Fig. \ref{Time_PAPR_BER_2s}(b), and it is evident that they are identical. Despite the fact that the suggested approach can transmit twice as much data as the typical OFDM system, it has the same BER performance, as illustrated in Fig. \ref{Time_PAPR_BER_2s}(c). 

\begin{figure}[!ht]
	\centering
	\includegraphics[width=0.95 \linewidth,height=12.0 cm]{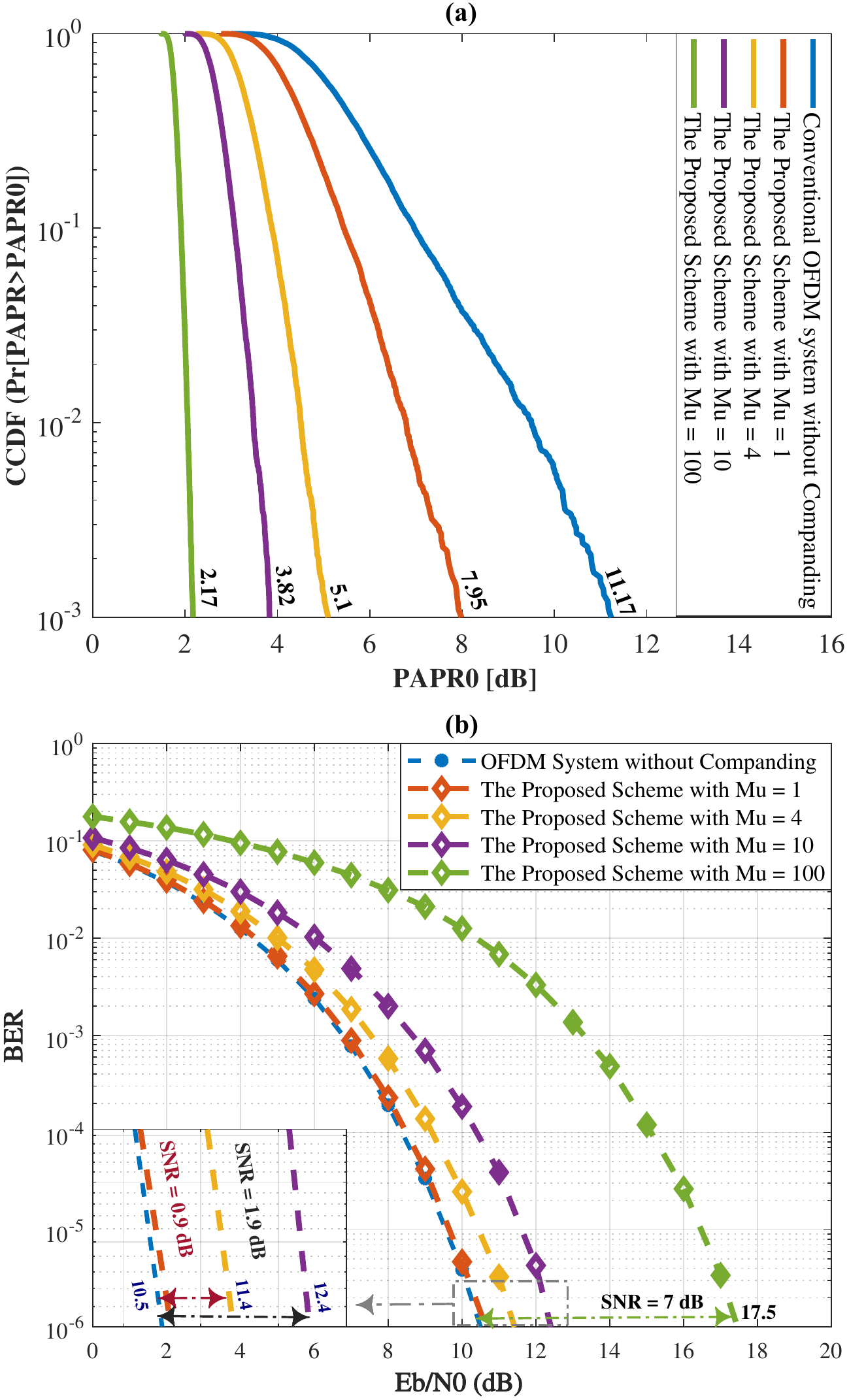}
	\caption{A typical OFDM system against the suggested method using $\mu$-Law companding techniques: (a) PAPR and (b) BER.}
	\label{PAPR_BER_CST_2S_Diff_Mu}
\end{figure}

The $\mu$-law companding approach reduces the high PAPR by enlarging small signals \cite{mohammed2019performance}. And therefore, it will be used with the proposed method  to enhance system performance and reduce PAPR. Fig. \ref{PAPR_BER_CST_2S_Diff_Mu} shows the PAPR and BER for the suggested method using  $\mu-$law companding technique. When the $\mu$ parameter is set to one ($\mu$ = 1), the PAPR improves by 3.22 dB with nearly no BER deterioration (BER degradation $\approx$ 0), as seen in Fig. \ref{PAPR_BER_CST_2S_Diff_Mu}(a) and (b). Increasing the value of $\mu$ parameter leads to decrease the PAPR and increase the improvement in PAPR, as displayed in  Fig. \ref{PAPR_BER_CST_2S_Diff_Mu}(a). However, this enhancement will be at the expense of degradation in BER, as it is clear in Fig. \ref{PAPR_BER_CST_2S_Diff_Mu}(b). Consequently, there is a trade off between the enhancement in PAPR and Degradation in BER. Table \ref{BER_PAPR_Table} summarizes the performance comparison of the proposed approach using $\mu-$law Companding and traditional OFDM based on PAPR and BER.

\begin{table}[!ht]
	\renewcommand{\arraystretch}{1.4}
	\centering
	\caption{PAPR improvement and BER deterioration for the suggested method using the Mu-Law technique.}
	\begin{tabular}{|c|c|c|}
		\hline
		$\mu-$\textbf{law} & \textbf{Improvement in PAPR} & \textbf{SNR at BER $10^{-6}$ degradation} \\
		\hline
		\hline
		$\mu$  = 1  & 11.17 - 7.95 = 3.22 dB & 10.5 - 10.5 = 0 dB  \\
		\hline
		$\mu$  = 4 &  11.17 - 5.1 = 6.07 dB & 11.4 - 10.5 = 0.9 dB  \\
		\hline
		$\mu$  = 10 &  11.17 - 3.28 = 7.89 dB & 12.4 - 10.5 = 1.9 dB  \\
		\hline
		$\mu$ = 100 & 11.17 - 2.17 = 9 dB & 17.5 - 10.5 = 7 dB  \\
		\hline
	\end{tabular}
	\label{BER_PAPR_Table}
\end{table}

\section{\textbf{Conclusion}}
The major issue of low power wide area networks (LPWAN) is supporting a large number of devices while employing a limited spectrum resources. This difficulty can be solved by employing narrow-band transmissions using NB-IoT. As a result, exponentially connected sensor nodes may be combined with additional benefits like better SNR and expanded coverage. Nonetheless, as the need for IoT services grows, more devices must be connected. The total number of linked devices is restricted because of the limited spectrum resources. The simulation results demonstrated that the STC-OFDM technique has the same performance as the standard OFDM system while conserving 50\% of bandwidth, using half the number of sub-carriers to transmit the same data as the conventional OFDM system. According to simulation studies, the STC-OFDM scheme decreases OFDM symbol time by half for the same amount of data when compared to the standard OFDM system. When employing the STC-OFDM approach, the number of IFFT is decreased by half, as is the PAPR, where the PAPR improved by 1.58 dB with zero degradation in BER and nearly the same complexity. In this study, the proposed method was employed to exploit the unused bandwidth  in order to double the number of connected devices without requiring more bandwidth while still maintaining the system performance. However, the suggested approach had the same PAPR and BER performance as conventional OFDM system. When compared STC-OFDM with Fast-OFDM, the two techniques had the same performance in sending the same amount of data with half the bandwidth (50\% of the bandwidth), as  compared to the typical OFDM. Additionally,  they had the same BER compared to the typical OFDM. The STC-OFDM, however, outperformed the Fast-OFDM by 1.35 dB in terms of the PAPR problem. To improve the system performance in the proposed method and lower PAPR, the $\mu$-law companding technique was combined with the proposed scheme. The $\mu$ value was carefully modified to get a good improvement in PAPR with no deterioration in BER. Based on the simulation results,  the $\mu$ parameter was set to one ($\mu$ = 1) to improve the  PAPR by 3.22 dB with almost no BER deterioration (BER degradation $\approx$ 0). In addition to the  benefits of reduced PAPR, improved system performance, and double the number of the connected devices, the proposed method using $\mu-$Law technique had roughly the same complexity as the conventional OFDM. Finally, raising the value of the $\mu$ parameter reduced PAPR and increased the improvement in PAPR. However, this improvement came at the cost of BER deterioration. As a result, there was a trade-off between the improvement in PAPR and the degradation in BER.


\bibliographystyle{unsrt}
\bibliography{./main}

\end{document}